\documentclass[10pt,twocolumn,letterpaper]{article}

\usepackage{wacv}
\usepackage{times}
\usepackage{epsfig}
\usepackage{graphicx}
\usepackage{amsmath}
\usepackage{amssymb}

\usepackage{subcaption}
\usepackage{amsmath,amssymb,amsfonts}
\usepackage{textcomp}
\usepackage{algorithm}
\usepackage{algpseudocode}
\usepackage{lmodern}
\usepackage{siunitx}
\usepackage{booktabs}
\usepackage{etoolbox}
\usepackage{array}
\newcolumntype{?}{!{\vrule width 1pt}}
\usepackage{mathtools}
\usepackage{multirow}
\usepackage{footmisc}
\usepackage{enumerate}

\def\wacvPaperID{10} 

\wacvfinalcopy 

\ifwacvfinal
\def\assignedStartPage{9876} 
\fi

\usepackage[accsupp]{axessibility}  

\ifwacvfinal
\usepackage[breaklinks=true,bookmarks=false]{hyperref}
\else
\usepackage[pagebackref=true,breaklinks=true,colorlinks,bookmarks=false]{hyperref}
\fi

\ifwacvfinal
\else
\fi

\usepackage{nopageno}

\begin{document}

\title{Powerful Physical Adversarial Examples Against Practical Face Recognition Systems}




\author{Inderjeet Singh\textsuperscript{1}\\
{\tt\small inderjeet78@nec.com}\\
\and
Toshinori Araki\textsuperscript{1}\\
{\tt\small toshinori\_araki@nec.com}\\
\and 
Kazuya Kakizaki\textsuperscript{1}\\
{\tt\small kazuya1210@nec.com}\\
\vspace{-3em}
\and
\textsuperscript{1}NEC Corporation\\
Kawasaki, Kanagawa, Japan\\
}






\maketitle

\begin{abstract}
It is well-known that the most existing machine learning (ML)-based safety-critical applications are vulnerable to carefully crafted input instances called adversarial examples (AXs). An adversary can conveniently attack these target systems from digital as well as physical worlds. This paper aims to the generation of robust physical AXs against face recognition systems. We present a novel smoothness loss function and a patch-noise combo attack for realizing powerful physical AXs. The smoothness loss interjects the concept of delayed constraints during the attack generation process, thereby causing better handling of optimization complexity and smoother AXs for the physical domain. The patch-noise combo attack combines patch noise and imperceptibly small noises from different distributions to generate powerful registration-based physical AXs. An extensive experimental analysis found that our smoothness loss results in robust and more transferable digital and physical AXs than the conventional techniques. Notably, our smoothness loss results in a 1.17 and 1.97 times better mean attack success rate (ASR) in physical white-box and black-box attacks, respectively. Our patch-noise combo attack furthers the performance gains and results in 2.39 and 4.74 times higher mean ASR than conventional technique in physical world white-box and black-box attacks, respectively.
\end{abstract}

\section{Introduction}
The adversarial machine learning (AML) domain has been expanding rapidly recently due to increased adversarial attacks on various traditional ML and deep learning (DL) systems. An adversarial attack is a process of causing well-planned misclassifications from a target classifier. In short, let us assume a ML system $f$, input sample $x_{clean}$, which is a natural input (Non-adversarial) and its true label $y_{true}$. The ML system correctly classifies this input i.e. $f(x_{clean})=y_{true}$. Now the system will be said under adversarial attack if an adversary carefully crafts an input sample $x_{adv}$ such that $f(x_{adv})\neq y_{true}$. 

An adversary can attack a target ML-based system from the digital world as well as the physical world. When the adversary leverages its digital access to the target system, creates and presents an adversarial example (AX) digitally to the target system, the realized attack is called a digital adversarial attack. On the other hand, it is called a physical adversarial attack if the adversary realizes the attack from the physical world. The target system for these AXs could be a face recognition system (FRS), where a trained DL model tries to validate the claimed identity of an input face image.

The adversarial attack can be of dodging type or impersonation type based on the attacker's objective. However, crafting impersonation (targeted) attacks is more challenging than dodging-type attacks. In this work, we focus on the generation of impersonation attacks. Also, an adversary can attack a target system during the training \cite{barreno2006can, biggio2011support, biggio2013poisoning, mei2015using}, testing \cite{goodfellow2014explaining, papernot2016distillation, papernot2016limitations, eykholt2018robust}, or model deployment stage\cite{miao2021machine, sehatbakhsh2020security}.

For physical adversarial attacks, the standard methods to transfer digital AXs to Physical World include printing and painting. The success of physical attacks depends on physical transferability, which is the ability of digital AXs to successfully transfer/imitate to the physical world. However, the attack performance in the physical world has been relatively lower than in the digital world. 

Some practical FRSs use a printed image to verify a subject’s identity, e.g., person re-identification, automatic ID document (like a passport) photo-matching systems at international borders. When an adversary submits a physical adversarial image in a registration process that causes a mistake by the target FRS, this attack is called a physical image registration attack. It allows an adversary to impersonate a target identity; This causes serious security concerns\cite{kakizaki2021toward}. This work considers physical image registration setting during physical world evaluation and focuses on improving physical transferability.

\begin{figure}
    \centering
    \includegraphics[width=\linewidth]{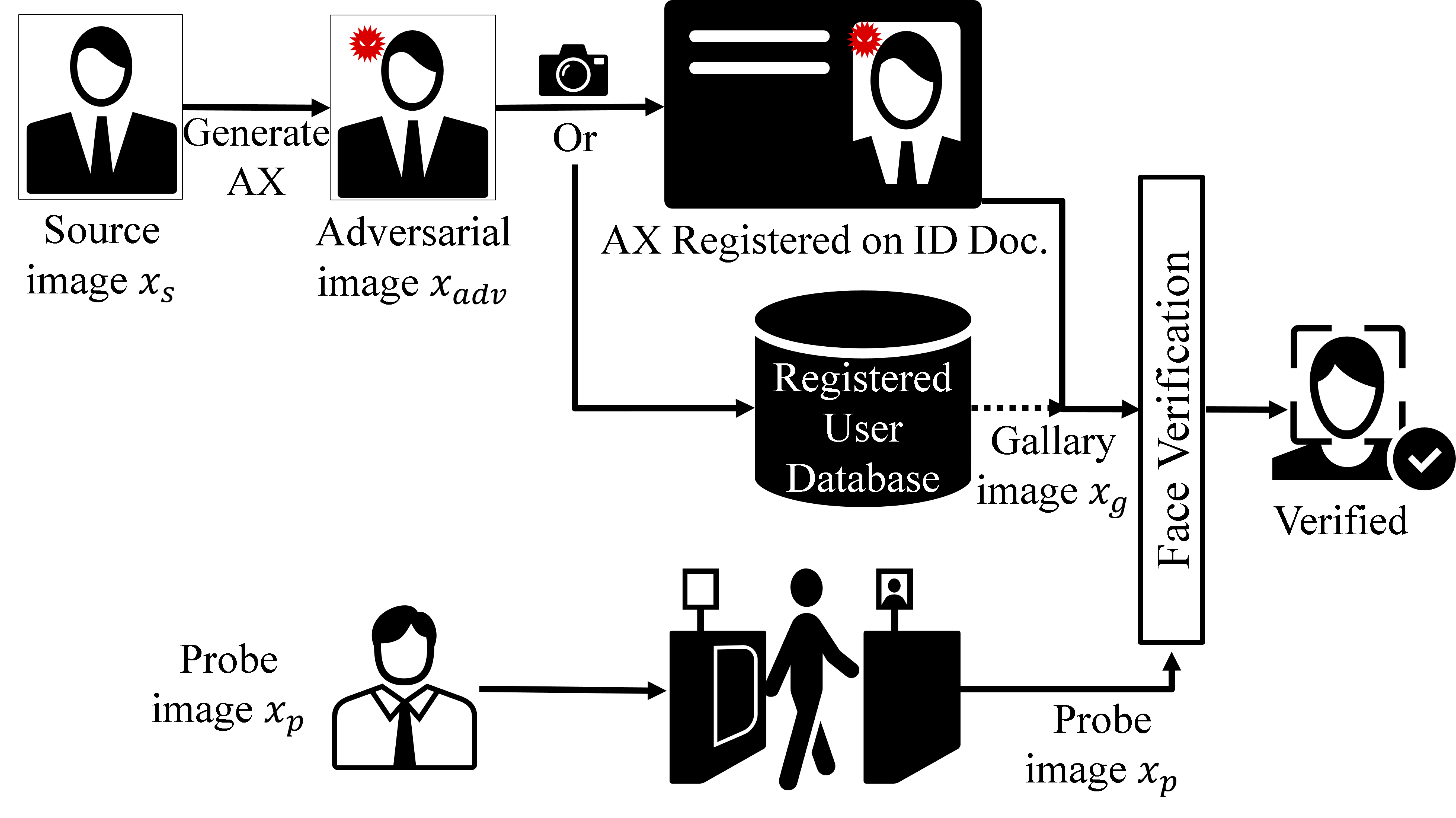}
    \caption{Physical image registration attacks against FRSs. An adversary registers an AX $x_{adv}$ generated using a clean source image $x_s$ in the target system.  At the time of face verification, the registered $x_{adv}$ results in the successful verification of a person at the security gate even the person ($x_p$) is different from the person of $x_{adv}$ image.}
    \label{fig:reg_attack}
\end{figure}

Effective optimization of adversarial noise during the attack generation process results in robust AXs following misclassification objectives. For a better physical attack success rate (ASR), the generation of smooth patterns is a well-known approach for lower \textit{physical reconstruction losses}. \textit{Physical reconstruction loss} is the total amount of information lost and the noise added to the physical AX while transferring it from the digital world\cite{kurakin2016adversarial,singh2021brightness}. The errors during printing, camera noises, change in camera angles, variable lighting conditions \cite{singh2021brightness}, attack surface characteristics, and realism is the critical parameters influencing the physical re-constructability of digital AXs, thereby the success of a physical AX.

\noindent \textbf{Our main contributions from this work are as follows:}
\begin{enumerate}
    \item We propose a novel smoothness regularizer for generating robust AXs with high physical transferability for attacking FRSs from the physical world. Our smoothness loss also provides black-box performance improvements to the generated attacks due to the regularization effect. 
    \item We propose a novel patch-noise combo attack method for generating powerful digital and physical adversarial attacks against FRSs by combining patch and imperceptibly small adversarial noises.
    \item We perform extensive white-box and black-box evaluation in the digital and physical worlds for state-of-the-art attack generation algorithms, for the proposed smoothness regularizer and the patch-noise combo attack method. We present a practical methodology for the physical world evaluation of registration-based adversarial attacks. We also provide the physical world evaluation for the ineffectiveness of the attacks with imperceptibly small noises in the physical world. 
\end{enumerate}

\subsection{Related Works}
To reduce physical reconstruction loss, previous works focused on generating smoother patterns in the adversarial noise. Existing studies found that abrupt pixel variations in a digital image, when compared to smoother patterns, cause significant printing \cite{sharif2016accessorize} and camera \cite{mahendran2015understanding} errors when transferred to the physical world. The smooth adversarial patterns result in a better ASR for the printed AXs. Sharif et al. \cite{sharif2016accessorize} proposed a smoothness loss function to generate smoother adversarial patterns for physical AXs. Their smoothness loss is given below.

\begin{equation}\label{eqn: tvm}
    T V(r)=\sum_{i, j}\left(\left(r_{i, j}-r_{i+1, j}\right)^{2}+\left(r_{i, j}-r_{i, j+1}\right)^{2}\right)^{\frac{1}{2}}
\end{equation}

\noindent Where $r_{i,j}$ is the pixel intensity in the noise image with coordinates $(i,j)$. This smoothness penalty is used as a regularizer in the adversarial loss function and minimizes the differences in the adjacent pixel values in an image $r$. They improved the smoothness of the adversarial patch, thereby improving the inconspicuousness and physical realizability of the generated adversarial patches. 

Though the smoothness loss used by Sharif et al. \cite{sharif2016accessorize} results in the performance of improvements of the generated AXs in the physical world, it suffers from some primary limitations. It causes more substantial constraints on the optimization procedure for adversarial attack generation, reducing the convergence rate and white box ASR in the digital environment. Also, this reduction in the feasible solution space restricts a large number of possible practical patterns. It treats the imperceptible noise as well as perceptible noise in the same manner. Moreover, it does not allow selective optimization of the adversarial noise. It does not allow the formation of smooth sub-patches within the main malicious patch when the pixel distribution within the sub-patch and the remaining area of the patch is substantially different, thus, making it unfeasible for generating AXs with a sub-region representing a small smooth area of significantly different colors or patterns.

\section{Our Proposed Smoothness Loss}
We propose a smoothness loss based on an activation threshold $\tau$ that causes calculation of total variation (TV) only for the pixel pairs having deviation greater than $\tau$ from an initial reference value. Our proposed smoothness penalty is given in equation \ref{our_loss}.
\begin{multline}\label{our_loss}
     L_{\text {smooth}}=\sum_{i, j}(\left(p_{i+1, j}-p_{i, j}\right)^2 \cdot\left(M_{i+1, j} \cdot M_{i,j}\right)+ \\ \left(p_{i, j+1}-p_{i, j}\right)^2 \cdot\left(M_{i, j+1} \cdot M_{i, j}\right))^{\frac{1}{2}}
\end{multline}

\noindent Where, 
\begin{equation}
M_{i, j}= \begin{cases}1 & p_{i, j} \geq \tau_{i,j} \text{ and } \tau_{i,j} \in Z \\ 0 & \text {else}\end{cases}
\end{equation}

\noindent $p_{i,j}$ is the difference in the value of a pixel from the reference value at location $(i,j)$ in the adversarial noise region. $M$ is a dynamic mask that activates the smoothness loss for pixels with intensities greater than a predefined level. The threshold matrix $Z$ activates the smoothness penalty for respective pixels in the adversarial patch region. The matrices $M$ and $Z$ have the exact dimensions as the source face image $x_s$ using which AX is generated. Each pixel-level threshold $\tau_{i,j} \in Z$ can have a predefined fixed value or can be updated iteratively.

\subsection{Working of Existing \cite{sharif2016accessorize} vs. Our Smoothness Loss}
The existing smoothness loss \cite{sharif2016accessorize} starts penalizing the total variation (TV) present in the input image's trainable pixels (adversarial noise pixels) right from the beginning of the attack generation process. This causes excessive constraints on the adversarial optimization process from the start, considerably limiting the feasible solution space for the adversarial noise patterns. On the other hand, our smoothness loss starts penalizing the TV only for pixel pairs with a significant deviation from an initial reference value. Thus the pixels away from the initial reference value only are smoothened. 

It is to be noted that giving very high weightage to our smoothness regularizer might result in confining the solution space near a ball (with radius proportional to thresholds in $Z$) around the reference value, but the constraints remain softer than the existing penalty \cite{sharif2016accessorize} because of the hinge type margin caused by $Z$. Hence our smoothness loss introduces delayed smoothness constraints only at selected pixel locations during the attack generation process resulting in faster and better converged final solutions while maintaining a similar level of smoothness present in the generated AXs.


\section{Our Patch-Noise Combo Attack}
Existing works focus on generating adversarial noise coming from a single distribution. To generate stronger AXs for FRSs, we combined an adversarial patch (eyeglass) noise $\delta_p$ with an imperceptibly small noise $\delta_s$ (calling imperceptible noise from hereon). The small noise $\delta_s$ is placed in the remaining area of the face image $x_s$.

\begin{equation}\label{patch-noise update}
    x^{p-n}_{adv} = x_s + M_s \cdot \delta_s + M_p \cdot \delta_p
\end{equation}
\noindent Where, the mask matrix $M_p$ takes values 1 for the pixels at patch location in $x_s$ and 0 otherwise. 

\begin{figure}
    \centering
    \includegraphics[width=\linewidth]{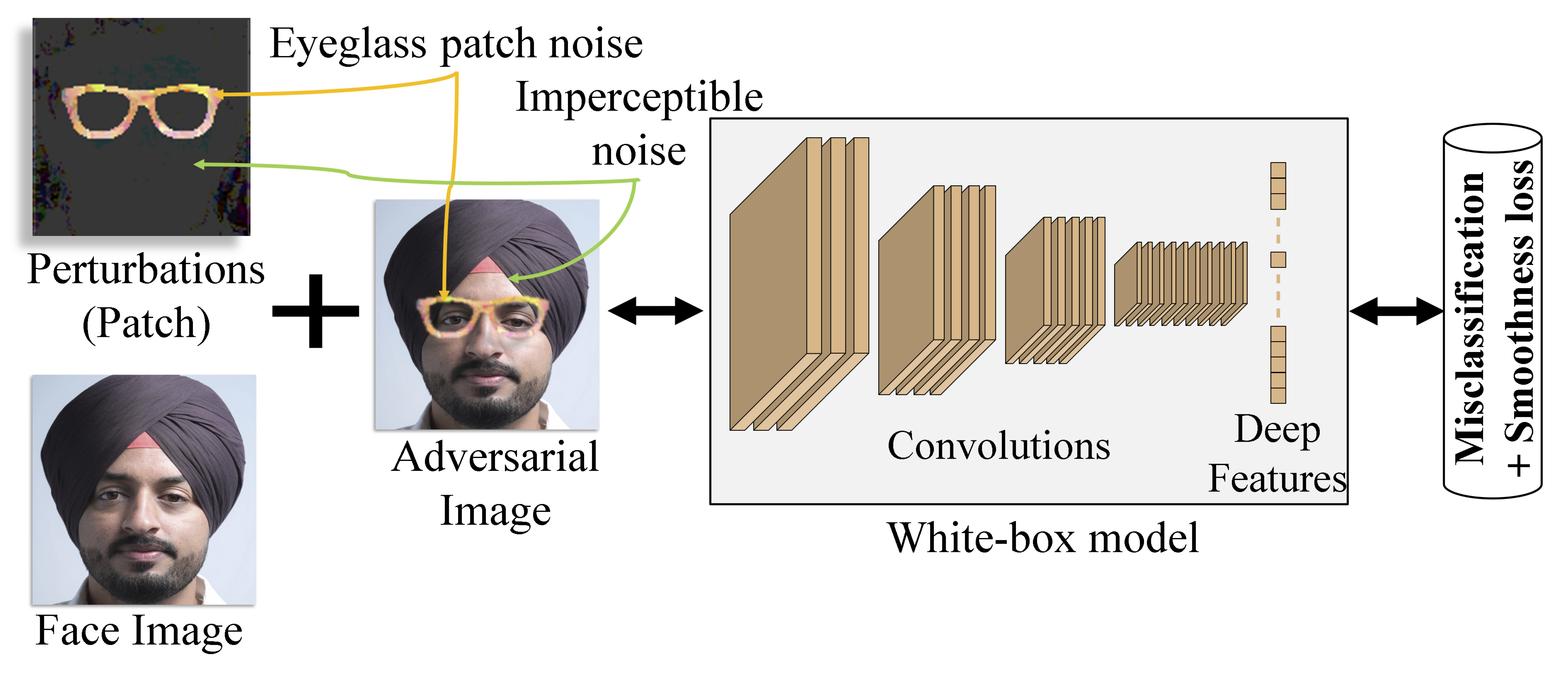}
    \caption{Generation of our patch-noise combo attack combining imperceptibly small noise with eyeglass patch for highly transferable digital and registration-based physical AXs.}
    \label{fig:noise_combo}
\end{figure}

In this attack, adversarial noises of different sizes and different distributions are combined to form a single strong AX. We abbreviate it as “patch-noise combo attack” due to the combination of adversarial noises coming from different distributions. Our patch-noise combo attack is illustrated in the Figure \ref{fig:noise_combo}. This attack causes increased ASR in the digital domain mainly because the increased feasible solution space resulted from additional adversarial noise. However, for the physical world success of this attack, the size constraints on the imperceptible noise play a vital role (can be understood further in Section \ref{res_dis}). 

The minimal size of the imperceptible noise $\delta_s$ results in digital world ASR increase only because the physical reconstruction losses neutralize the effect in the physical world. However, increasing the size (by relaxing $L_{\infty}$ size constraints) increases physical ASR but decreases the physical imperceptibility of generated AXs. Hence, a target domain-specific choice of the size constraints for $\delta_s$ can be made for generating powerful patch-noise combo attacks in the physical world.

This type of attack can be successfully used for attacking FRSs digitally and physically where the submission of a printed image is required for the verification, e.g., registration attacks.

\section{Experimental Setting} \label{exp_setting}
\subsection{Pretrained DL Models}
We used state-of-the-art DL models as deep feature extractors in the FRS setting. In the feature extractor setting, the final layer of a DL model outputs a vector of deep representative features for the given input compared to the output class probabilities in the classification setting. A properly trained classifier can be used as a feature extractor by removing the last softmax/classification layer.

We used ResNet50\cite{he2016deep}, ResNet100\cite{he2016deep}, VGG16\cite{vgg16model}, SE-IR100\cite{hu2018squeeze}, and SphereFace\cite{Liu_2017_CVPR} model architectures. The VGG16 model was trained on VggFace\cite{vggface_data} data using the softmax loss. We trained two variants for each of ResNet50 and ResNet100 models using softmax and arcface\cite{deng2019arcface} loss functions on VggFace2\cite{vggface2} and MS1MV2\cite{deng2019arcface} data, respectively. The IR-SE100 model was also trained on MS1MV2 dataset. We calculated the verification threshold $\tau$ for best accuracies. Only four models were used at a time for each attack setting. The diverse model setting allows the elimination of unnecessary biases during the evaluation.

\subsection{Algorithms for Attack Generation}
For a comprehensive experimental analysis, we selected the four most widely used gradient-based attack generation algorithms. We applied all these algorithms in a feature-extractor setting. The algorithms that we use for the attack generation are given below.   

\noindent \textbf{(1) Projected Gradient Descent (PGD) Attack \cite{pgd}.} We initialized the PGD attack for source image $x_s$ with some initial Gaussian random noise $\delta$ as $x_0 \leftarrow x + \delta$.  The update step during $t^{th}$ iteration of attack generation process for the PGD attack is given in equation \ref{pgd_update}.

\begin{equation}\label{pgd_update}
    x^{t+1} \leftarrow {Clip}_{x,\epsilon}\left(x^t + \alpha \cdot Sign\left(\nabla L_{adv}\right)\right)
\end{equation}
\noindent Where, $\nabla L_{adv}$ is the gradients calculated w.r.t the adversarial loss function $L_{adv}$.

\noindent \textbf{(2) Carlini and Wagner Attack (CW) \cite{cw}.} The Carlini Wagner attack minimized the $\Delta L$ and training adversarial noise $\delta$ while keeping $x+\delta \in [0,1]$ for all pixels in the adversarial region of the training image. For making sure to satisfy the above constraints, the following transformation was used.
\begin{equation}
    x_i + \delta_i = \frac{tanh(w_i)+1}{2}
\end{equation}
Where, $w_i$'s were trainable parameters.

\noindent \textbf{(3) Layerwise Origin-Target Synthesis (LOTS) Attack \cite{lots}.} This attack is proposed explicitly for feature extractors. In our setting, we calculated the loss for the final layer of the model only.

\noindent \textbf{(4) Iterative Fast Gradient Sign (IFGSM) Attack \cite{ifgsm1,ifgsm2}.} This attack is almost the same as the PGD attack except the training images were initialized as $x_0 \leftarrow x$.


\subsection{Black-box Attack Techniques}
To generate black-box attacks, we used three techniques for attack robustness and transferability enhancement.

\noindent \textbf{(a) Input Diversity Method \cite{input_div}:} we applied random crop transformations to the training image in each training iteration of the attack generation process. The random crop was limited to $7\%$ of the original image's length for every four edges.

\noindent \textbf{(b) Ensemble Diversity Method \cite{ensemble_div}:} we used ResNet50\cite{he2016deep} and SE-IR50\cite{hu2018squeeze} models (with equal weights to loss) out of the four pre-trained DL models for attack generation and remaining models for black-box evaluation.

\noindent \textbf{(c) Combination of Input and Ensemble Diversity: }we combined the input diversity \cite{input_div} and ensemble diversity \cite{ensemble_div} to further enhance the robustness and black-box transferability.

\subsection{Type of Attacks and Adversarial Objective}
All of the experimentations were performed for an impersonation attack objective. In the impersonation attacks against FRSs, a source image $x_s$ is perturbed to an adversarial image $x_{adv}$ considering an adversarial loss for impersonation $L_{imper}$, such that the deep features of $x_{adv}$ becomes similar to the deep features of a target identity's image $x_{tar}$. Hence, the adversarial image $x_{adv}$ is classified as $x_{tar}$ by the target model, causing a mistake by the target DL model. We used the following loss functions for generating impersonation attacks:
\begin{equation}\label{adv_loss_fn}
    {L}_{adv}= \gamma \cdot {L}_{smooth} - \sum_{i}^{K} f_{d}\left(f_{i}\left(x_{t}\right), f_{i}\left(x_{train} \right)\right)
\end{equation}

\noindent Where, ${L}_{smooth}$ is either the TV loss or our smoothness loss function, $x_{train}$ is the face image $x_s$ combined with adversarial perturbations. For patch only attacks, $x_{train} = x_s + M_p \cdot \delta_p $. For the patch-noise combo attacks, $x_{train}$ follows equation \ref{patch-noise update}. $x_t$ is the image of target identity. $f_i$ is the pre-trained face feature extractor. For non-ensemble models, $K = 1$ and $f1$ is one of the feature extractors. For ensemble models \cite{ensemble_div}, $K=2$, and $f_1$ and $f_2$ represents ResNet50\cite{he2016deep}, SE-IR50\cite{hu2018squeeze} feature-extractors. $\gamma$ is the weight parameter for the smoothness loss function. For the instances with input diversity transformation \cite{input_div}, the training image become $x^{'}_{train} = f_{DI}\left( x_{train} \right)$. The function $f_{DI}$ applies input diversity transformation \cite{input_div}. $f_d$ calculates the $L_2$ distance between given inputs.

\subsection{Ablation Study Instances}

\begin{algorithm}
\caption{Attack Evaluation Procedure Outline}\label{alg:attack_gen_procedure}
\begin{algorithmic}
\State \hskip-0.6em \textbf{Inputs:} Source image $x_s$, target image $x_t$; algorithms A = \{PGD,CW,LOTS,IFGSM\}; black-box methods B = \{None, DI, Ensemble, DI+Ensemble\}; attack techniques T = \{No regularizer, TV Loss, Our Loss, Patch-Noise Combo+TV Loss, Patch-Noise Combo+Our Loss\};
\State \hskip-0.6em \textbf{For} all $a_i \in A$, $b_i \in B$, $t_i \in T$ \textbf{do}
\State Generate digital AXs $(X_{adv})$ using combination of $a_i$,$b_i$, and $t_i$, while following the loss function of equation \ref{adv_loss_fn} and update rule of $a_i$;
\State \hskip-0.6em \textbf{Perform} \textit{digital} evaluation of $X_{adv}$;
\State \hskip-0.6em \textbf{Perform} \textit{physical} evaluation of successful digital $X_{adv}$ as per section \ref{phy_pipeline};
\end{algorithmic}
\end{algorithm}

For the ablation study, we implemented a total of 80 attack combinations following Algorithm \ref{alg:attack_gen_procedure}; out of them, 48 were baselines, and 32 were our method’s instances. We took training face images for attack generation from VggFace2 data\cite{vggface2}. We generated 100 AXs for the digital evaluation for each attack combination and a subset of 10 AXs for the physical evaluation.

\subsection{Parameter Settings}
The attack parameters for all the baselines and our methods were kept identical for effective comparison between baselines and our methods.
We set training iterations to 2000 for PGD, LOTS, and IFGSM attacks and 7000 for CW attacks after observing the convergence for the given iterations. The $\epsilon$ parameter was kept to 1 for all attacks as we are generating patch attacks. The learning rate parameter was also kept identical for all instances at 0.01 in the gradient descent setting.

\begin{figure*}
    \centering
    \includegraphics[width=0.9\textwidth]{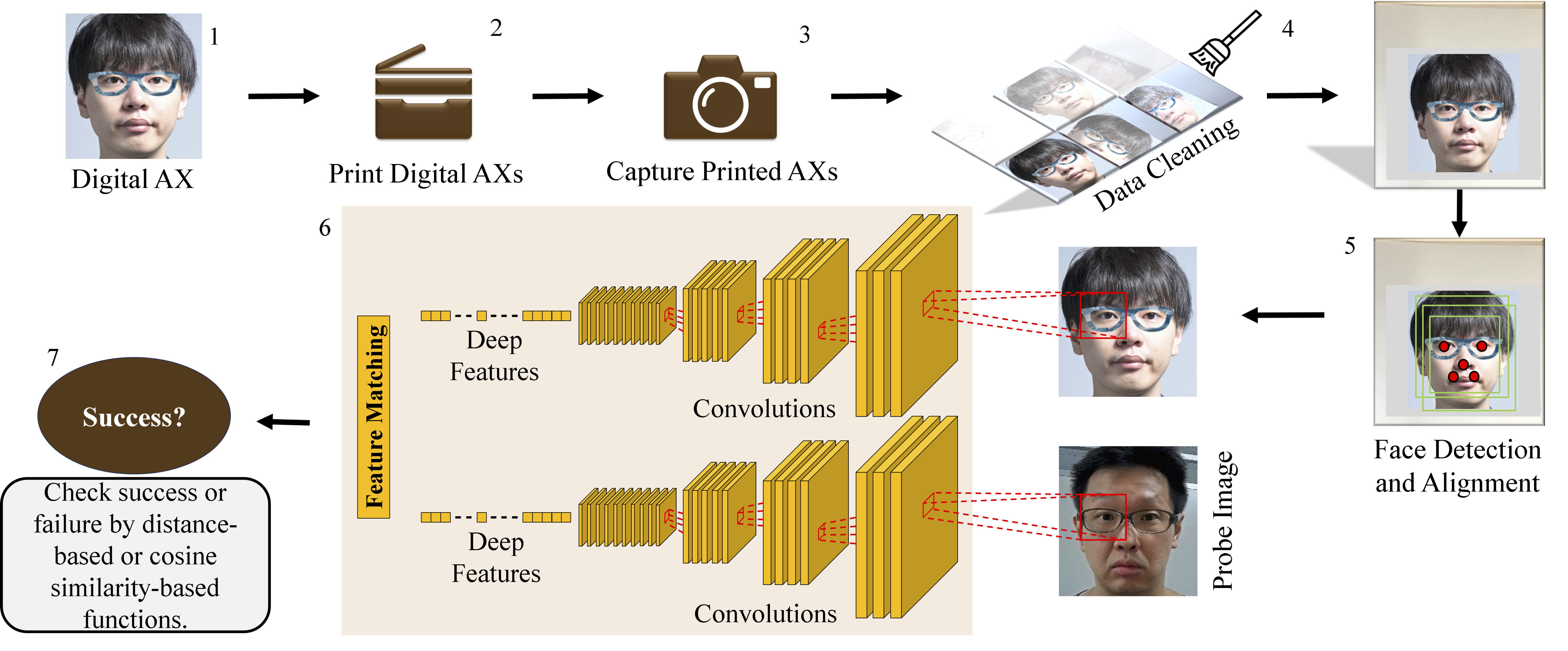}
    \caption{Our physical AX generation and evaluation pipeline in the registration attack setting.}
    \label{fig:phy_pipe}
\end{figure*}

\section{Physical AXs Generation Pipeline} \label{phy_pipeline}
We considered the light conditions and the camera angle, two main physical parameters during the realization of the physical AXs.

\noindent \textbf{1. Light conditions:} Two sub-parameters were considered for the light conditions parameter. \textbf{i. Brightness levels:} We considered two different brightness levels of 800 and 1200 lux. \textbf{ii. Light color temperature:} We captured the images under white light with color temperatures of 3000K, 5000K.

\noindent \textbf{2. Angle of the camera w.r.t. the printed adversarial image:} To capture the effect of reflectivity of the attack surface, we captured a stream of images by moving the camera in a horizontal arc of radius 15cm (approx.) and by subtending an angle of approximately 45° at the center of the captured image.

To transfer the successful digital AXs to the physical world for realizing physical adversarial attacks, we followed the pipeline shown in Figure \ref{fig:phy_pipe}. The steps are as follows:

\noindent\textbf{Step 1: Generating successful digital AXs.} Powerful physically transferable digital AXs were first generated. 

\noindent\textbf{Step 2: Printing.} The generated AXs were then color printed on paper.

\noindent\textbf{Step 3: Capturing printed AXs.} The printed AXs were then captured using a camera in an appropriate format for the target system. This step transfers the AXs in physical form back to the digital space for the evaluation.

\noindent\textbf{Step 4: Data cleaning.} From captured images, blurry and improper images were then cleaned. Around 20 images for each AX then remained.

\noindent\textbf{Step 5: MTCNN face detection \cite{zhang2016joint} and alignment.} To align and crop the captured physical AX as shown in Figure \ref{fig:phy_pipe} (after step-4), firstly, we performed MTCNN face detection. The detected faces were then aligned using the similarity alignment.

\noindent\textbf{Step 6: Feeding cropped and aligned AXs to the face matcher.} The cropped and aligned AXs were fed to a target face matcher depending on the attack evaluation setting (black/white-box).

\noindent\textbf{Step 7: Checking attack success.}Finally, the success or failure of the attack is checked by distance or cosine similarity-based thresholds obtained for best f1-scores.

\begin{figure*}
\centering
\begin{subfigure}{.5\textwidth}
  \centering
  \includegraphics[width=0.99\linewidth]{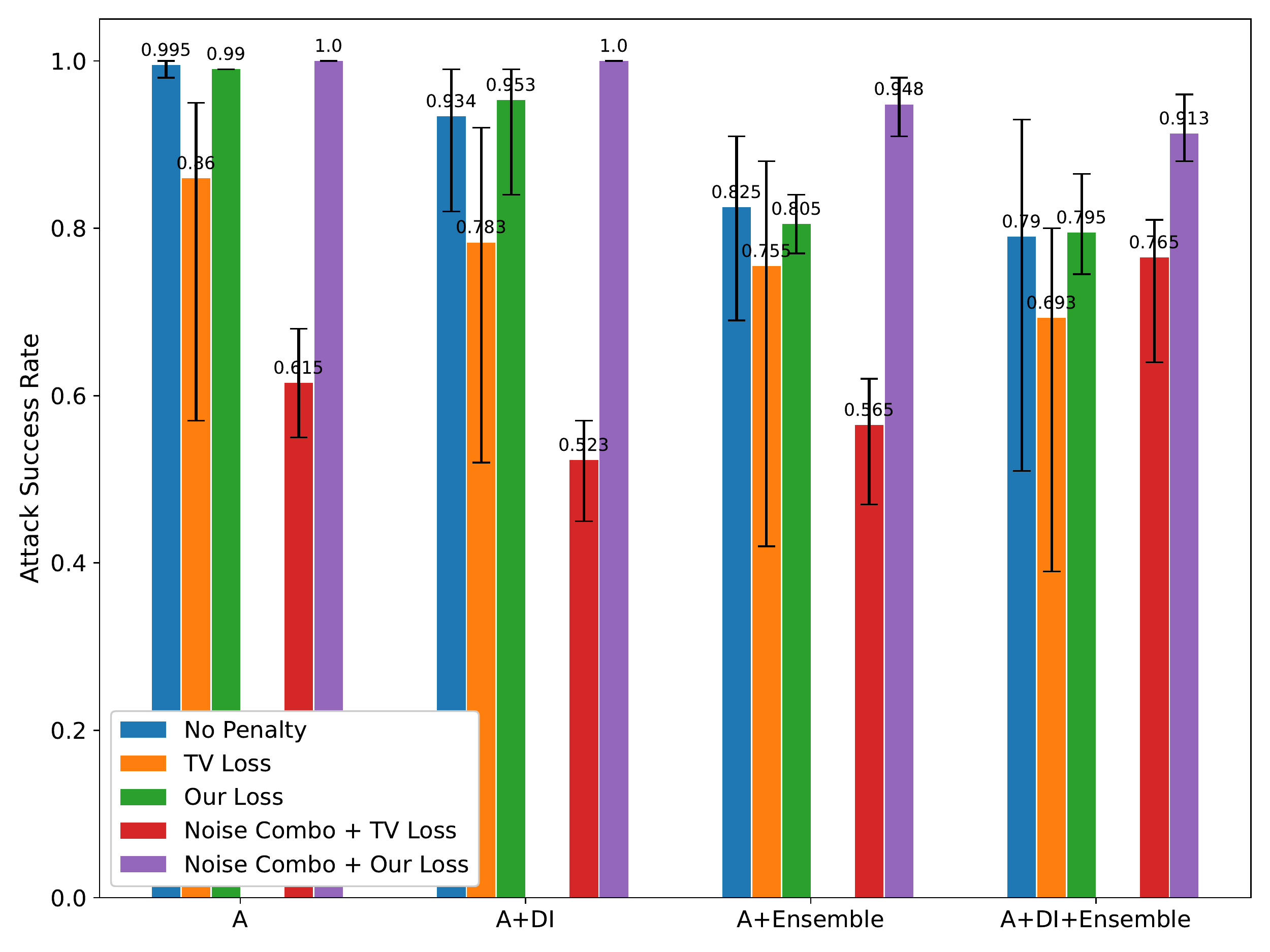}
  \caption{ASR for the white-box digital AXs}
  \label{fig:dg_wb}
\end{subfigure}%
\begin{subfigure}{.5\textwidth}
  \centering
  \includegraphics[width=0.99\linewidth]{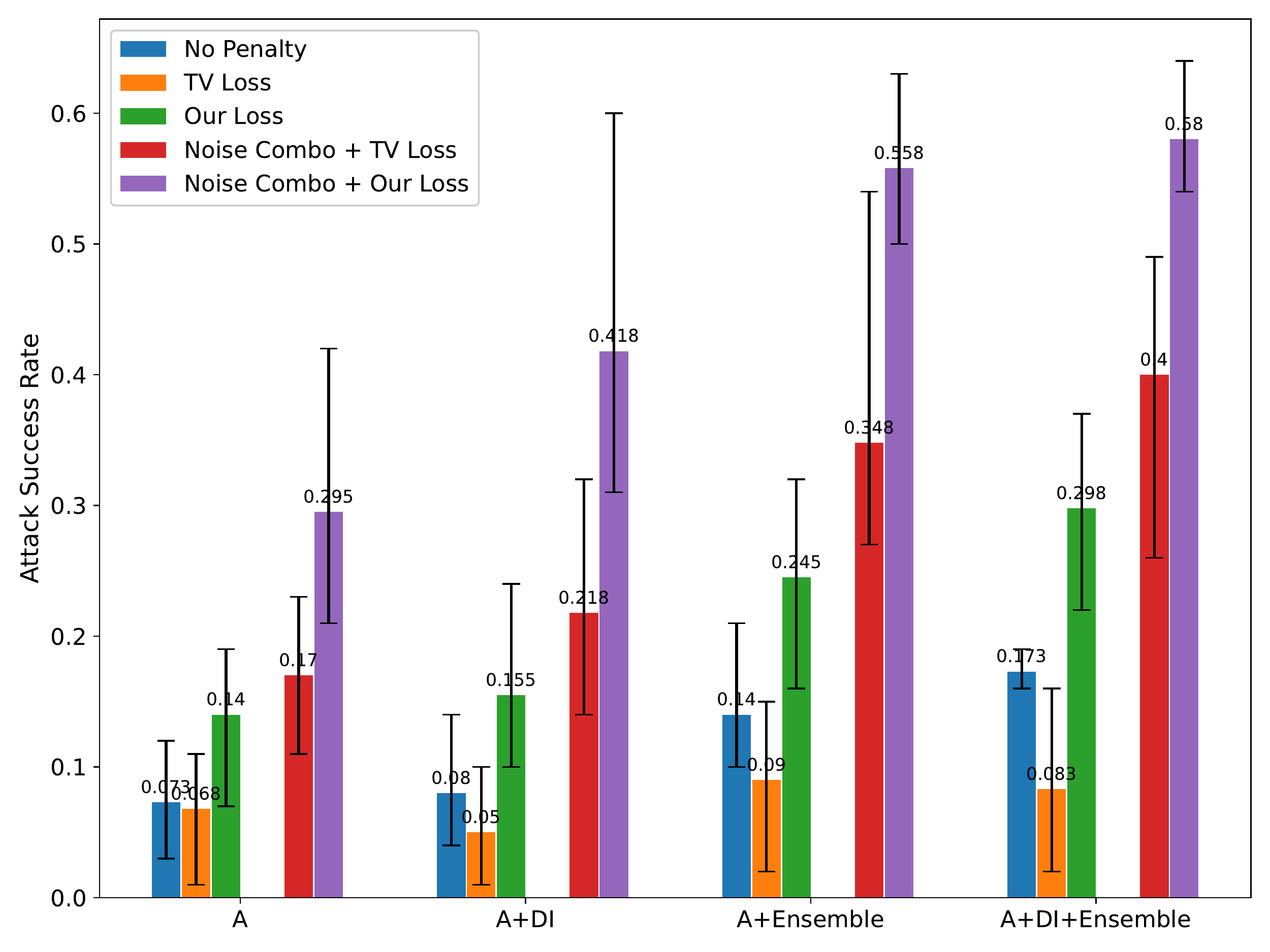}
  \caption{ASR for the black-box digital AXs}
  \label{fig:dg_bb}
\end{subfigure}
\caption{Mean adversarial ASR for white-box (left) and black-box (right) attacks in the \textit{digital} domain. The x-axis represents the different attack generation methods, and the y-axis represents ASR. The attack generation algorithms are denoted by ‘A’ and follow the combinations of Algorithm \ref{alg:attack_gen_procedure}.}
\label{fig:dig_results}
\end{figure*}

\section{Results and Discussion} \label{res_dis}
The comprehensive experimental analysis found that our smoothness loss and patch-noise combo attack method significantly exceeds conventional techniques, especially in the physical world. The black-box transferability was also found to be increased for the generated attacks. The performance gains over the TVM approach \cite{sharif2016accessorize} for both digital and physical worlds are mainly due to better optimization in the digital world, better handling of the areas to be smoothened in the adversarial patch, and convergence to more effective solutions. Our patch-noise combo attacks in the face recognition setting generated the most potent digital and physical domains attacks.

For the white-box attacks in the digital domain, we can see from Figure \ref{fig:dg_wb} that the ASR for AXs from our smoothness loss is almost the same as the AXs generated from without any smoothness regularizer. Ideally, the AXs generated without any smoothness regularizer for the white-box case should have higher ASR as they are trained for only that single adversarial objective without any additional constraints (regularizers). Nevertheless, the attacks generated with TVM regularizer had low ASR. The reason for low ASR was the stricter smoothness constraints on the attack generation procedure by calculating total variation right from the start for all pixels in the adversarial noise region. Our smoothness loss causes fewer constraints by the calculation of total variation during the attack generation process. The fewer constraints are due to the threshold-based activation of total variation calculation. Also, due to this, the AXs generated from our patch-noise combo attack method (using our smoothness loss) considerably outperform the ones generated from the TVM approach. The ASR for our patch-noise combo attack was 1.48 times higher than the eyeglass patch-only attacks because the more significant number of trainable parameters resulted in a significantly larger feasible solution space.

From the results for the digital black box attacks shown in Figure \ref{fig:dg_bb}, we found that the AXs generated using our smoothness loss significantly outperforms in the black-box transferability and resulted in 2.88 and 1.80 times higher mean ASR compared to the AXs generated using TVM and without TVM, respectively. Our patch-noise combo attack performed the best and recorded 6.36 times higher black-box mean ASR than TVM-based patch attack. However, the AXs generated from the patch-noise combo attack using TVM regularizer also had significantly better black-box performance than the other baselines. A good white-box ASR and not being overfitted are the two critical parameters that contribute to the better black-box performance of the adversarial noise.

\begin{figure*}
\centering
\begin{subfigure}{.5\textwidth}
  \centering
  \includegraphics[width=\linewidth]{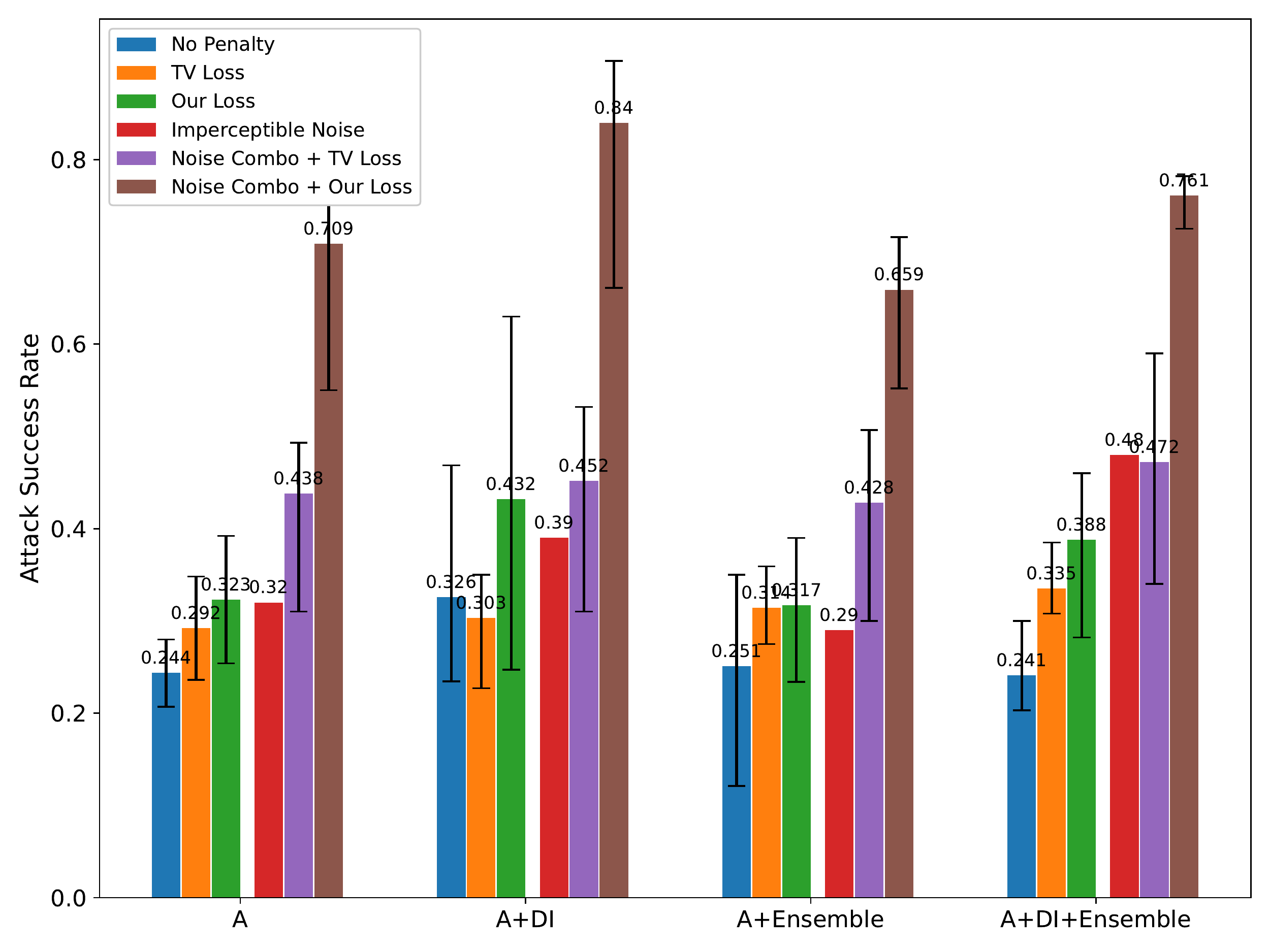}
  \caption{ASR for the white-box physical AXs}
  \label{fig:sub1}
\end{subfigure}%
\begin{subfigure}{.5\textwidth}
  \centering
  \includegraphics[width=\linewidth]{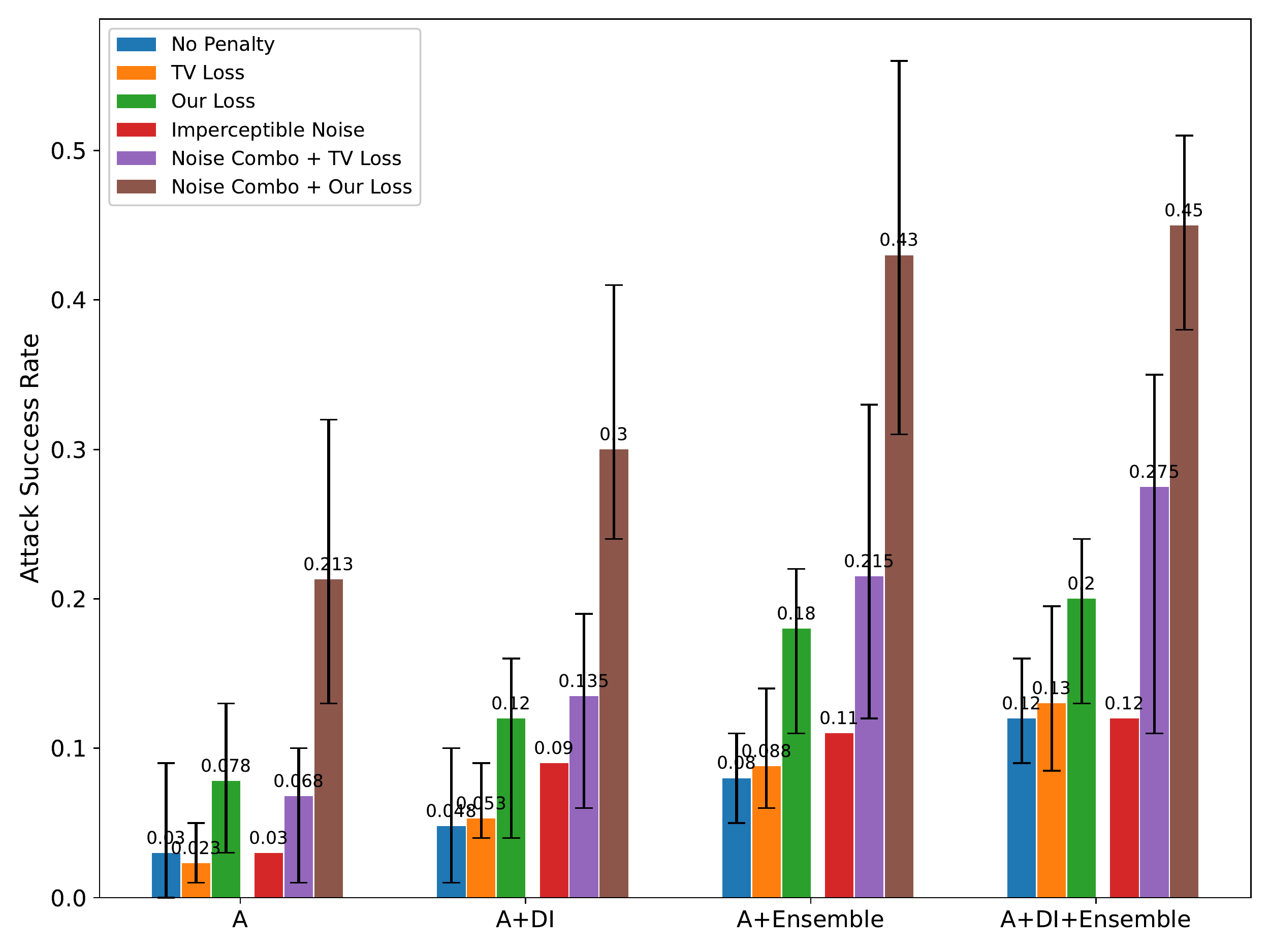}
  \caption{ASR for the black-box physical AXs}
  \label{fig:sub2}
\end{subfigure}
\caption{Mean adversarial ASR for white-box (left) and black-box (right) attacks in the \textit{physical} domain. The x-axis represents the different attack generation methods, and the y-axis represents ASR. The attack generation algorithms are denoted by ‘A’ and follow the combinations of Algorithm \ref{alg:attack_gen_procedure}.}
\label{fig:phy_results}
\end{figure*}

In the physical domain, the AXs from our smoothness loss and the patch-noise combo attack using our smoothness loss completely outperforms their respective baselines (Figure \ref{fig:phy_results}). The patch-noise combo attacks with our smoothness regularizer were the strongest in the physical domain and resulted in 2.39 and 4.74 times higher mean ASR for the white-box and black-box settings. The performance gap between ours and baseline methods in the physical domain was significantly higher than in the digital domain. Hence, proving the effectiveness of our smoothness loss and the patch-noise combo attack for excellent physical transferability of generated AXs.

\subsection{Comparing Smoothness in Generated AXs}
To better understand the superior performance of our smoothness regularizer and to confirm our hypothesis, we compare the total variation present in the generated AXs from ours and the conventional \cite{sharif2016accessorize} TVM-based smoothness regularizers.  
We calculated the total variation present in the AXs generated from the existing \cite{sharif2016accessorize} and our smoothness loss using the following equation \ref{eqn: tvm}.

We found that AXs generated without using any smoothness regularizer, using TVM regularizer \cite{sharif2016accessorize}, using our smoothness loss, using patch-noise combo attack with TVM regularizer, and using patch-noise combo attack with our smoothness regularizer had a mean total variation of 39.31, 19.39, 20.93, 20.25, and 22.75, respectively. 

We can see a more significant reduction in the total variation in the generated AXs due to a smoothness regularizer. But we can also see that this difference is not very significant compared to the smoothness of generated AXs from our and the existing TVM-based smoothness regularizer. Hence, we can say that our smoothness loss introduces a similar amount of smoothness in the generated AXs while allowing them to reach much better optimal solutions.

\subsection{Physical Transferability of Successful Digital AXs}

\begin{table}
    \centering
    \sisetup{detect-weight,mode=text}
    \renewrobustcmd{\bfseries}{\fontseries{b}\selectfont}
    \renewrobustcmd{\boldmath}{}
    \newrobustcmd{\B}{\bfseries}
    \addtolength{\tabcolsep}{0.29pt}
     \begin{tabular}{c|c?c|c|c|c}
     \toprule
     Sr.     &  Smoothness   & \multicolumn{4}{c}{ AX Generation Method } \\ \cline{3-6}
     No.     &  Regularizer             & A1                & A2                & A3                & A4        \\
     \midrule
          1      & S0      & $0.240$   & $0.340 $   & $0.320 $   &  0.315       \\
     \midrule
     2          & S1       & $0.250 $   & $0.345 $   & $0.447 $   &  0.379     \\
     \midrule
     3            & S2      &  $0.320 $ & $0.432 $   & $0.390 $   &  0.420       \\
     \midrule
         4       & S3       & $0.700 $   & $0.780 $   & $0.750 $   &  0.830    \\
     \midrule
         5       & S4       & $0.830 $   & $0.890 $   & $0.710 $   &  0.838    \\
     \bottomrule
     \end{tabular}
     \caption{Results for the Physical Transferability of the Successful Digital AXs. S0, S1, S2, S3, and S4 represents the use of no smoothness regularizer, existing smoothness loss \cite{sharif2016accessorize}, our smoothness loss, patch-noise combo and existing smoothness loss, and patch-noise combo and our smoothness loss.}
      \label{tab:physical_trans}
   \end{table}

Table \ref{tab:physical_trans} shows the physical transferability of the successful digital AXs to the physical domain. It can be seen that the AXs generated using any smoothness-based regularizer have better ASR in the physical domain as compared to the AXs generated without any smoothness regularizer. Also, the physical transferability of AXs generated from our smoothness loss remained better than the AXs generated from existing TV minimization even after using softer regularizing smoothness constraints.

\subsection{Imperceptible Noise Attacks in Physical World}
We also evaluate the performance of physical AXs with imperceptibly small adversarial noise distributed over the entire face image. Conventionally, these kinds of attacks are prevalent in the digital world only. However, in this work, we check whether they are usable for the physical world attacks. 

In our experimentation, we first generated digital attacks for 7 different values (0.02, 0.05, 0.1, 0.25, 0.5, 0.75, 1) of the $epsilon$ parameter. It is to be noted that source images were normalized in the range $[0,1]$. For each value of the $epsilon$ parameter, we generated three successful digital AXs with different identities for the physical evaluation. We transferred all digital AXs to the physical world and checked the physical ASRs.

\begin{figure}[!h]
    \centering
    \includegraphics[width=0.95\linewidth]{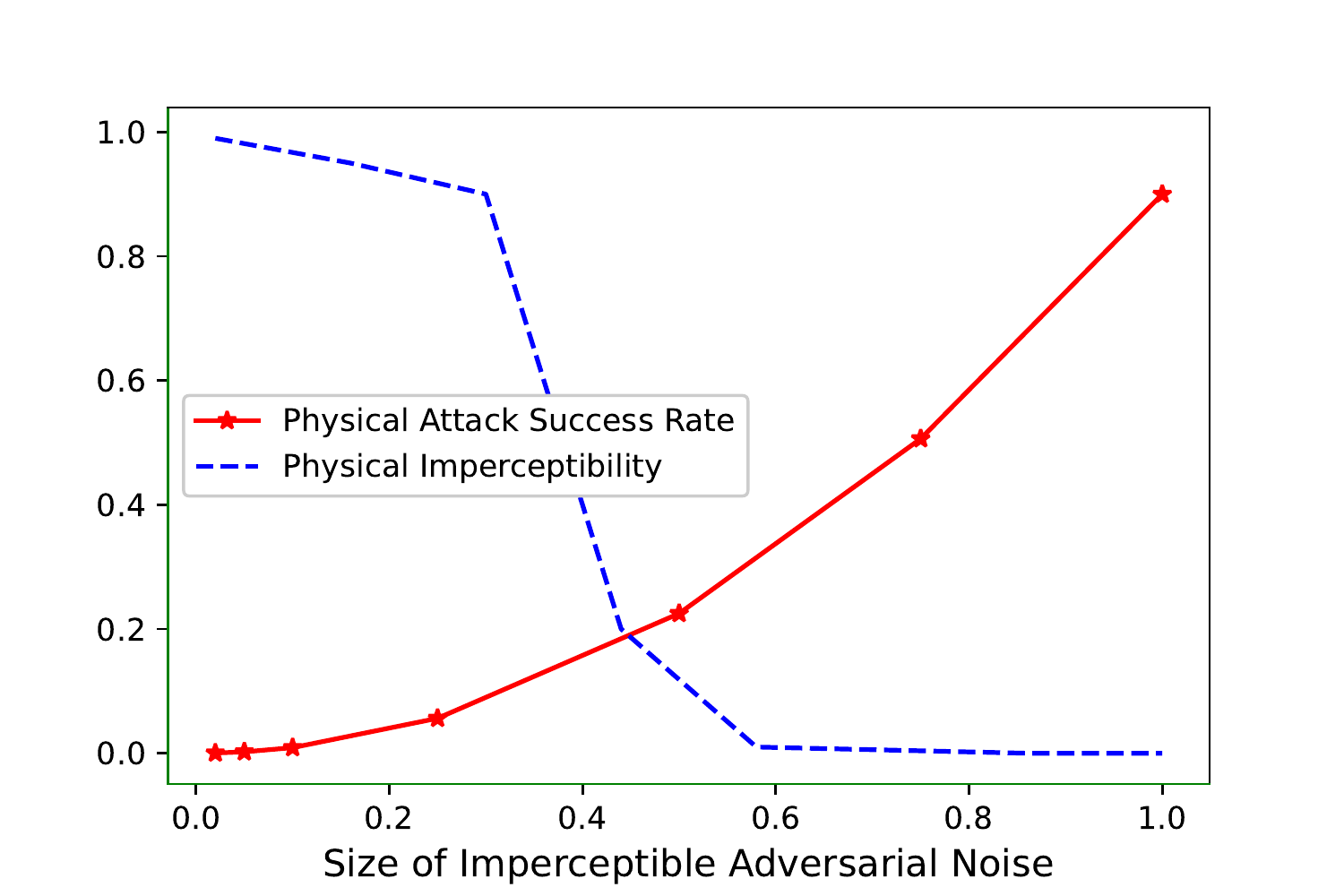}
    \caption{Change in physical ASR and imperceptibility with increasing size $\epsilon$ of the adversarial noise. Increasing $\epsilon$ increases physical ASR but decreases physical imperceptibility.}
    \label{fig:imper_size}
\end{figure}

From the results (Figure \ref{fig:imper_size}) for the white-box ResNet50\cite{he2016deep} model, we found direct proportionality between $\epsilon$ and physical ASR; This is because AXs with smaller $\epsilon$ have less prominent visible adversarial patterns to the target system. Less prominence or visibility is caused due to physical reconstruction losses. Also, the AXs with smaller $\epsilon$ are more sensitive to the small perturbations than those with larger $\epsilon$. Also, from a subjective evaluation, we found an inverse proportionality between $\epsilon$ and physical inconspicuousness.

Hence, we conclude a trade-off between the ASR and inconspicuousness of AXs in the physical domain. For a physical attack to succeed, it must be physically inconspicuous and have a high attack success probability. That makes the simple adversarial noise attacks (adversarial noise not taking any real-world patch shape) an unpopular choice for the physical adversarial attacks.

\section{Conclusions} \label{conclusions}
This paper proposed a novel smoothness regularizer and patch-noise combo attack method for generating powerful physical adversarial examples against practical FRSs. From extensive experimentation, we found that our proposed methods outperform all state-of-the-art baselines in the digital and physical worlds. 

The use of our smoothness loss results in better physical transferability. Our smoothness loss also allows generating much complex real-world adversarial patterns due to selective smoothening in adversarial noise, reducing unwanted boundary losses. In our patch-noise combo attack, we confirm that using imperceptibly small adversarial noise along with adversarial patches can result in \textit{significant} performance improvements in the physical world. However, the size of imperceptible noise has the trade-off of ASR and physical imperceptibility. The generated digital and physical AXs from our methods allow an adequate risk assessment for the practical FRSs.

{\small
\bibliographystyle{ieee_fullname}
\bibliography{main}
}

\end{document}